# Multiwavelength Passively Mode Locked Fiber Laser: all normal dispersion

**H. Zhang, D. Y. Tang\*, L. M. Zhao, and X. Wu**

School of Electrical and Electronic Engineering,

Nanyang Technological University, Singapore 639798

\*: edytang@ntu.edu.sg, corresponding author.

We report on the generation of multiwavelength gain-guided soliton (GGS) in an all normal dispersion birefringent fiber laser passively mode-locked with a semiconductor saturable absorber mirror (SESAM). Determined by the cavity birefringence, we observe the stable single-, dual- and triple- wavelength mode locked GGSs, which is a natural consequence of the mutual nonlinear interaction among the normal cavity dispersion, cavity fiber nonlinear Kerr effect, laser gain saturation and bandwidth limitation resulting from the intra-cavity birefringent filter.

Multiwavelength mode locked fiber laser has versatile applications including fiber optic sensing, instrumentation, photonic component characterization and especially WDM optical communication. Several approaches have successfully achieved multiwavelength mode locked lasing. Li et al reported the generation of triple wavelength picosecond mode locked pulses through a self-seeded Fabry–Perot laser diode with fiber Bragg gratings [1]. A multiwavelength actively mode-locked fiber laser incorporating a single sampled fiber bragg grating [2] or a biased semiconductor optical amplifier [3] has been proposed by Yao et al. By virtue of nonlinear polarization rotation (NPR) effect, simultaneous dual- and five-wavelength actively mode-locked erbium-doped fiber laser at 10 GHz has been realized by Pan et al [4]. Although multiwavelength actively mode-locked fiber laser has advantages such as high repetition rates, narrow linewidth and large amounts wavelengths, we could not obviate its drawbacks including: broad pulse width, weak peak power, cost inefficient and high insertion loss because of modulator incorporated in the cavity. Moreover, actively mode locked multiwavelength pulse with weak nonlinearity is impossibly shaped to be optical soliton which possess born preponderance: well stability, low time jittering, short pulse width and high peak power. However, if passively instead of actively mode locking technique is adopted, multiwavelength soliton tend to be formed. Back to 1992, Matsas et al had reported the experimental observation of dual-wavelength soliton exploiting NPR technique in a fiber laser made of purely negative dispersion [5].

Triggered by Matsas's experiment, we design an all normal dispersion fiber laser passively mode locked by SESAM to investigate the generation of multiwavelength soliton. Recently, formation of GGS in net normal dispersion fiber laser mode locked by

NPR [6], SESAM [7] and even a hybrid technique [8] were reported. In this paper, although the soliton spectral bandwidth is far narrower than the Erbium gain bandwidth, GGS could be still formed as a result of mutual nonlinear interaction among the normal cavity dispersion, cavity fiber nonlinear Kerr effect, laser gain saturation and filter bandwidth limitation due to the presence of an artificial intra-cavity birefringence filter. Furthermore, as the filter behavior depends on the cavity birefringence, tunable dual- and triple- GGS could be achieved. Particularly, in view of the fact that filter effect commonly exists in Yb-doped fiber laser cavity [9] where filter bandwidth limitation is much easier to be fulfilled than gain bandwidth limitation, GGS could be still formed and even multiwavelength GGS could be observed through refining the filter quality. We believe the findings not only fill the gap in obtaining multiwavelength passively mode locked solitons in normal dispersion regime but also generalized the theory of GGS to the other nonlinear system where filter bandwidth limitation effect dominates over gain bandwidth limitation.

Our fiber laser is schematically shown in Fig. 1a. The laser cavity is an all-normal dispersion fiber ring consisting of 5.0 m, 2880 ppm Erbium-doped fiber (EDF) with group velocity dispersion (GVD) of -32 (ps/nm)/km, and 8.5 m dispersion compensation fiber (DCF) with GVD of -4 (ps/nm)/km. Mode-locking of the laser is achieved with a SESAM. A polarization independent circulator was used to force the unidirectional operation of the ring and simultaneously incorporate the SESAM in the cavity. Note that within one cavity round-trip the pulse propagates twice in the DCF between the circulator and the SESAM. A 50% fiber coupler was used to output the signal, and the laser was pumped by a high power Fiber Raman Laser source (KPS-BT2-RFL-1480-60-FA) of wavelength 1480 nm. The maximum pump power can be as high as 5W. All the passive components used (WDM, Coupler, and Circulator) were made of the DCF. An optical spectrum analyzer (Ando AQ-6315B) and a 350MHZ oscilloscope (Agilen 54641A) together with a 2GHZ photo-detector were used to simultaneously monitor the spectra and pulse train, respectively. The SESAM used is made based on GaInNAs quantum wells. It has a saturable absorption modulation depth of 30%, a saturation fluence of 90 µJ/cm$^2$ and a recovery time of 10 ps. It is specially designed together with 0.5m DCF pigtails to ensure the absence of negative dispersion fiber in the cavity. Fig. 1b shows the measured low-intensity reflectivity of the SESAM. The high absorption (low reflectivity) was found in the wavelength of 1570nm and the absorption spectrum is not wavelength isotropic but wavelength dependent. It means that SESAM could function as a bandwidth filter influencing the laser operation and facilitating the multiwavelength generation.

This fiber laser is an all normal dispersion ring cavity with positive cavity GVD of about 0.2522 ps$^2$. Experimentally, once the mode locking self starts, we could obtain a typical GGS spectrum with steep spectral edges, similar to the observation of GGS in a dispersion managed cavity with net positive dispersion mode locked through SESAM [7]. As soliton generation and propagation in this fiber laser cavity are strongly cavity birefringence dependent, the output soliton features could be varied significantly through rotating the paddles of polarization controllers (PCs). Adjusting PCs, central wavelength of GGS could be varied in a wide range from 1570nm to 1590nm. But the formed GGSs with central wavelengths close to 1570 or 1590nm are less stable because of the detuning from the optimum operation window of laser around 1585nm where EDF has the maximum gain co-efficiency and SESAM has an appropriate saturable absorption. However, if the pump strength is increased till above 300mw and simultaneously fine tune PCs, except a steep square spectrum oscillating at one wavelength, another steep square spectrum lasing at the other wavelength could be observed, as shown in Fig. 2a, which is actually dual wavelength GGS mode locking. Moreover, if PCs is slightly shifted from the current position, GGS at one wavelength is suppressed while the other GGS is boosted. Then rotating PCs in an opposite direction results in a reversed variation and such repeatable process could sustain over a long time. In order to clarify the formation mechanism of dual wavelength GGSs, its corresponding autocorrelation trace and oscilloscope trace are monitored. Fig. 2b shows that two groups of GGSs with different soliton energies, represented by two different pulse heights in the oscilloscope, propagate with different group velocities in the cavity, demonstrating each GGS in the time domain corresponds to GGS centered at one wavelength in the frequency domain.

Further polarization resolved experiments reveal that one group of GGS is suppressed while the other GGS is still maintained through rotating the external cavity polarizer, from both the optical spectrum and oscilloscope trace, i.e., GGS centered at 1576.2nm correlates with the pulse train with higher intensity, GGS centered at 1579.6nm relates to the pulse train with lower intensity and their polarization are orthogonal to each other. It thus indicates that the mode-locked pulses are not only dual wavelengths but also dual polarizations. Because these two group GGSs collide with each other endlessly and therefore their time separations are flexible, only one symmetric pulse profile within the autocorrelation measurement range is found and it is impossible to determine the pulse width for individual GGS, as shown in the insert of Fig. 2a. It has a width (FWHM) of ∼ 48ps. If a $sech^2$ pulse shape is assumed, the soliton pulse width is ~28.8 ps. The 3dB spectrum bandwidth of the soliton centered at 1576.2nm (1579.6nm) is ∼ 3.4nm (3nm), which gives a time-bandwidth product about 12.2 (10.8), indicating that both GGSs are still strongly chirped.

Starting from the current state, slowing reducing the pump strength, only spectral intensity becomes weaker but the spectral bandwidth, central wavelength and pulse duration for each GGS are nearly kept constant. Once the pump power is lower than 110 mw, the output jumps from the original dual wavelength GGS to single wavelength GGS. Surprisingly, the central wavelength of single wavelength GGS could be either 1576nm or 1579nm, i.e, these two group GGSs shown in Fig.2 are physically equivalent in stability and which one could survive at low pump depends on bifurcation routes. If the pump strength is further reduced till mode locking is completely restrained, two continuous wave peaks located at 1576 nm and 1579 nm are observed, indicating the existence of intra-cavity birefringence filter with transmission maxima located at 1576 nm and 1579 nm. Afterwards, single GGS and dual GGS could recur subsequently if pump strength is gradually increased.

Based on the above experimental observation, the mechanism of dual wavelength GGSs can be explained as follows. Despite of the fact that the core/cladding diameters of DCF spliced in the cavity are nearly equal to that of normal dispersion EDF, unavoidable artificial birefringent filter, resulting from moderate birefringent EDF and misalignments of principle axes between EDF and DCF, should be taken in into consideration, especially under strong cavity birefringence and nonlinear birefringence at high pump strength. EDF then acts as a Mach-Zehnder interferometer, with its principal axes acting as interferometer arms. It is widely known that the wavelength spacing between the Mach-Zehnder transmission maxima is $\Delta\lambda = \lambda^2/LB$, where $\lambda$ is the central wavelength, $L$ is the EDF length and $B$ is the EDF's degree of birefringence. The measured wavelength spacing of dual wavelength GGS is about 4nm, indicating the EDF degree of birefringence of $1.24 \times 10^{-4}$, which is comparable to that reported in polarization maintaining fiber [10]. In the presence of intra-cavity filter, GGS could be still formed as a result of the mutual nonlinear interaction among the normal cavity dispersion, cavity fiber nonlinear Kerr effect, laser gain saturation and bandwidth limitation. In contrast with previously investigated GGS where bandwidth limitation arises from the EDF gain bandwidth limitation, the filter bandwidth limitation dominates over the gain bandwidth limitation. It can be inferred that the applicable scope of GGS is not restricted in the fiber laser system described by the Ginzburg-Landau equations, where gain bandwidth limitation effect could not be ignored, but general in other nonlinear system systems governed by the Swift-Hohenberg equation, where filter effect should be included [11]. Although around each transmission maxima of the filter, individual GGS could be formed separately within the filter bandwidth, they are not isolated. As in the fiber laser, the laser gain is polarization independent. After the formation of a GGS with horizontal polarization centered at one wavelength, most of the laser gain which supports the generation of this GGS with horizontal polarization can be depleted as a result of the laser gain competition. However, the laser gain which sustains the formation of soliton with vertical polarization is still in effect. Consequently, the newly generated GGS at the other wavelength could potentially have larger chance to be vertically polarized. This can explain why the observed dual GGSs are orthogonally polarized.

In the experiments, although the experimental conditions for the generation of dual wavelength GGSs is rigor in that the cavity birefringence and pump power must be kept in an appropriate range, it could be very stable over a long period as a whole day. As intra-cavity filter performance could be varied through controlling the cavity birefringence, central wavelength, amplitudes, spectrum bandwidths and wavelength separation of the dual wavelength GGS could be tuned in a wide range. We even observe the transformation of dual wavelength GGS to single wavelength GGS, as shown in Fig. 3a, through rotating PCs but the pump strength is kept fixed. If simultaneously monitored by an oscilloscope trace, we find that the previously observed two group GGSs merge together and finally one GGS exists in the cavity. No fine structures are observed through a high speed oscilloscope (50GHz) combined with a commercial autocorrelator (FR-103MN). Regarding dual wavelength GGSs, their group velocities are different owing to such large difference in central frequencies, it is impossible to observe the phenomena of incoherent trapping of dual wavelength GGSs but they always collide with each other in the time domain. However, tuning the PCs could balance the cavity birefringence and modify the filter performance, eventually; these two groups of GGSs coincide and co-propagate together as one unit, as shown in Fig. 3b. Different from dual wavelength GGSs whose two polarization components are nearly linearly and orthogonally polarized, the current single wavelength GGS is elliptically polarized, i.e., its two orthogonal components have comparable spectral intensity and the same central wavelength but clearly different spectral distributions [7]. Therefore, the two orthogonally polarization components are coherently coupled and could be termed as gain guided vector soliton. Thanks to the strong and efficient coherent coupling forces which could compensate both cavity birefringence and nonlinear birefringence, the two polarization components could be trapped together in the time domain and polarization locked in the frequency domain despite of strong frequency chirps. The phase locking between the two orthogonal polarizations was confirmed by the polarization evolution frequency measurement as described in [12-15]. After passing through an external cavity polarizer, the pulse height on the oscilloscope trace would have identical value as shown in Fig. 3b.

Besides dual wavelength GGS, triple wavelength GGS, as shown in Fig.5, could be obtained if the cavity birefringence is appropriately set. It is also a product of the intra-cavity birefringent filter. Once the PCs are finely tuned in order to generate such filter with three transmission maxima within the EDF gain profile, triple wavelength GGS with equal wavelength separation around ~4nm is observed.Contrast with dual wavelength GGSs, triple wavelength GGSs are more sensitive to the external perturbations.

In conclusion, we have proposed a multiwavelength GGS fiber laser and investigated its generation mechanism. We find that an artificial birefringence filter exists in the cavity and its performance could be regulated through alternating the cavity birefringence. Accordingly, stable single-, dual-, and triple- wavelength mode locked GGSs are successively obtained. Moreover, their frequencies, amplitudes, spectrum bandwidths and wavelength spacing are tunable in a wide range.

**Figure captions:**

Fig. 1: (a) Schematic of the experimental setup. EDF: Erbium doped fiber. WDM: wavelength division multiplexer. DCF: dispersion compensation fiber. PC: polarization controllers; (b) Normalized reflectivity of the SESAM measured at low-intensity input.

Fig. 2: (a) Optical spectra of dual wavelength GGSs. Insert: the corresponding autocorrelation trace and normalized optical spectrum; (b) oscilloscope trace of dual wavelength GGSs.

Fig. 3: (a) Optical spectra of polarization locked gain guided vector soliton and dual wavelength spectrum obtained through rotating PCs but kept the pump strength fixed. (b) Oscilloscope trace of gain guided vector soliton after passing through a polarizer.

Fig. 4: Single/dual/triple wavelength spectra obtained through rotating PCs but kept the pump strength fixed. (a) dB units; (b) normalized units.

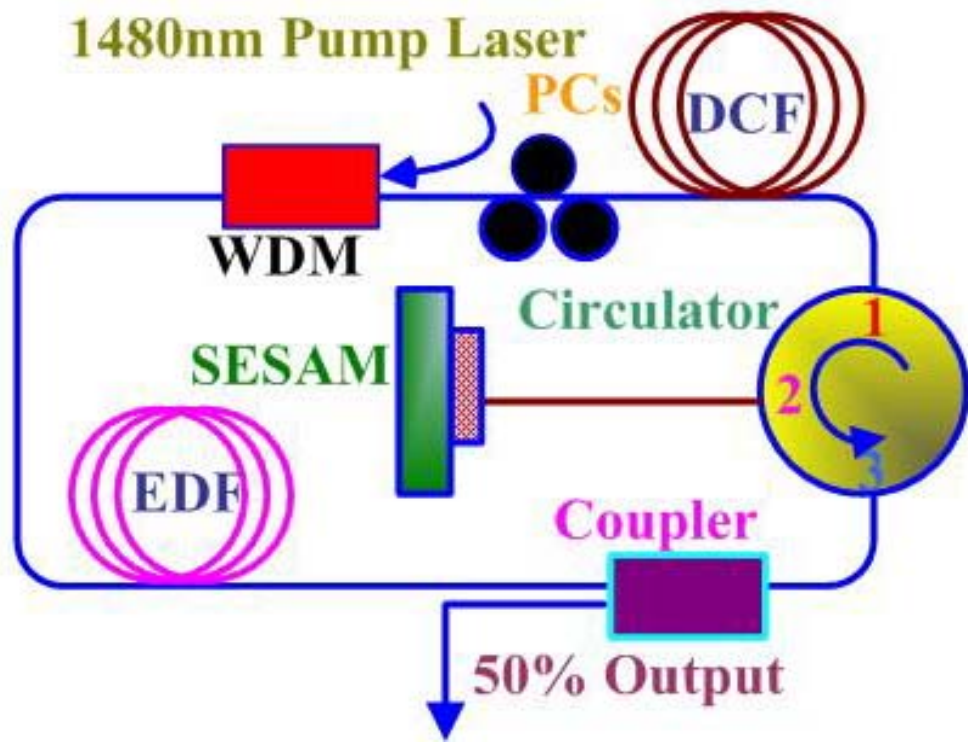

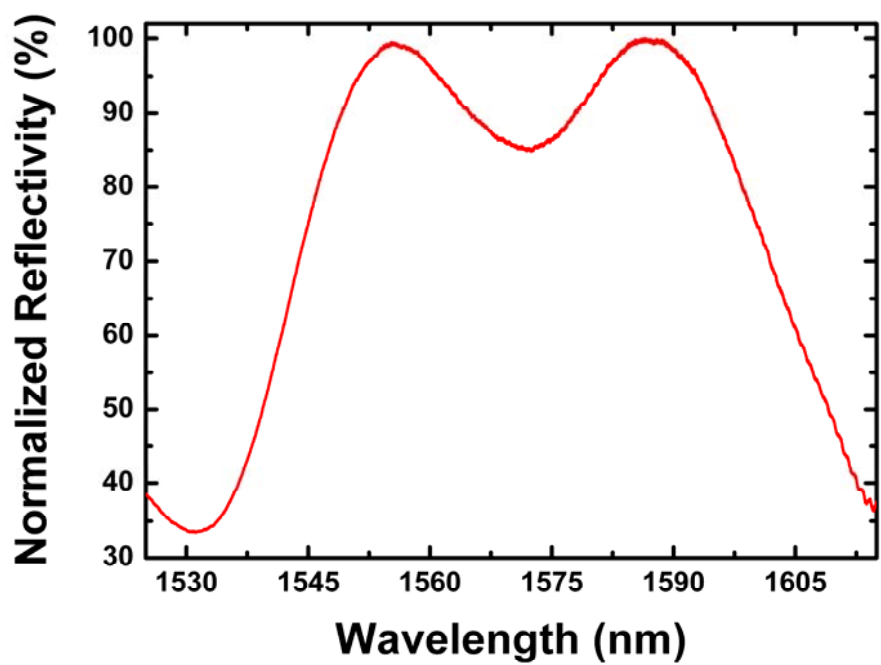

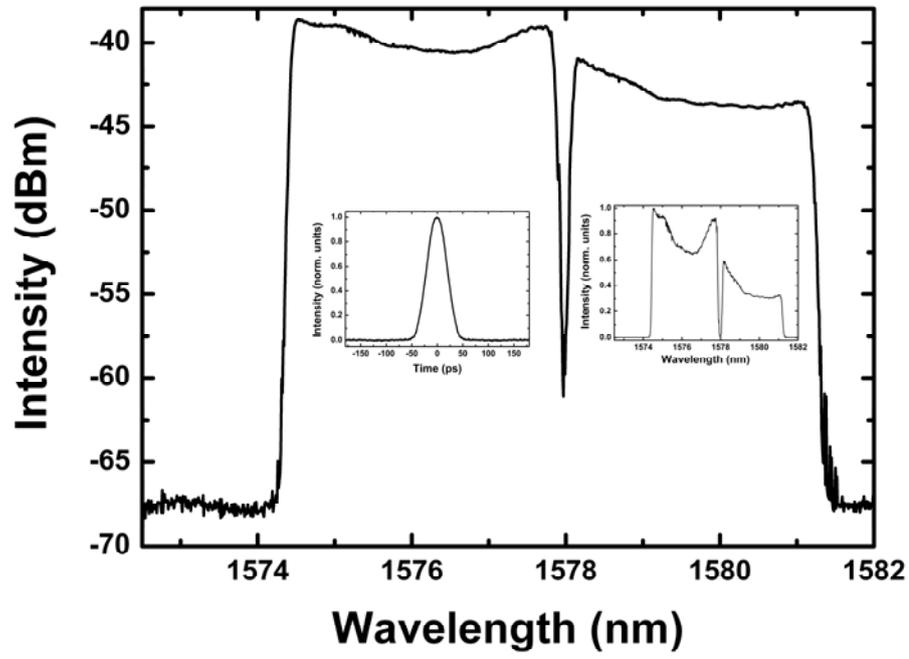

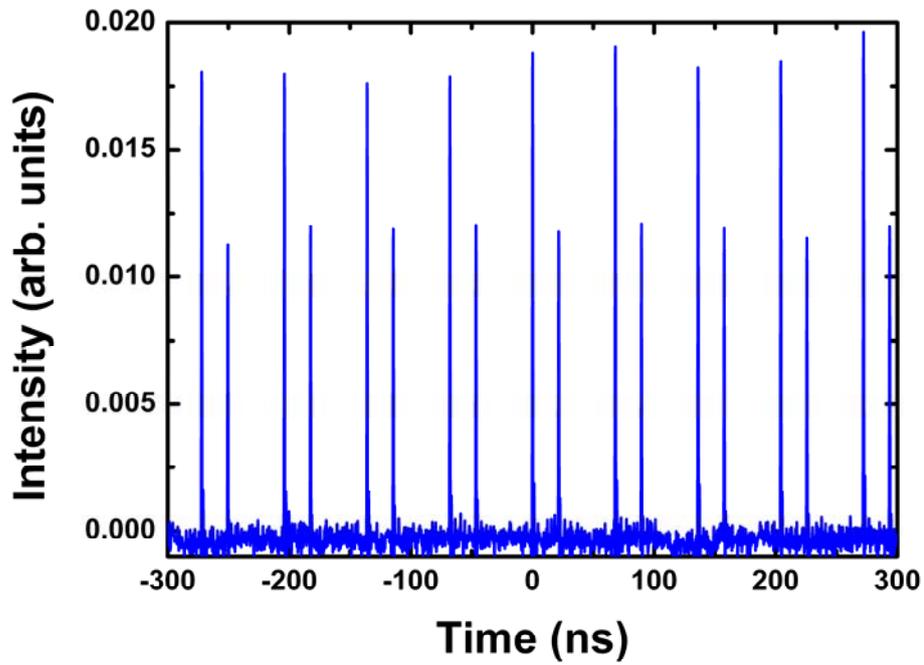

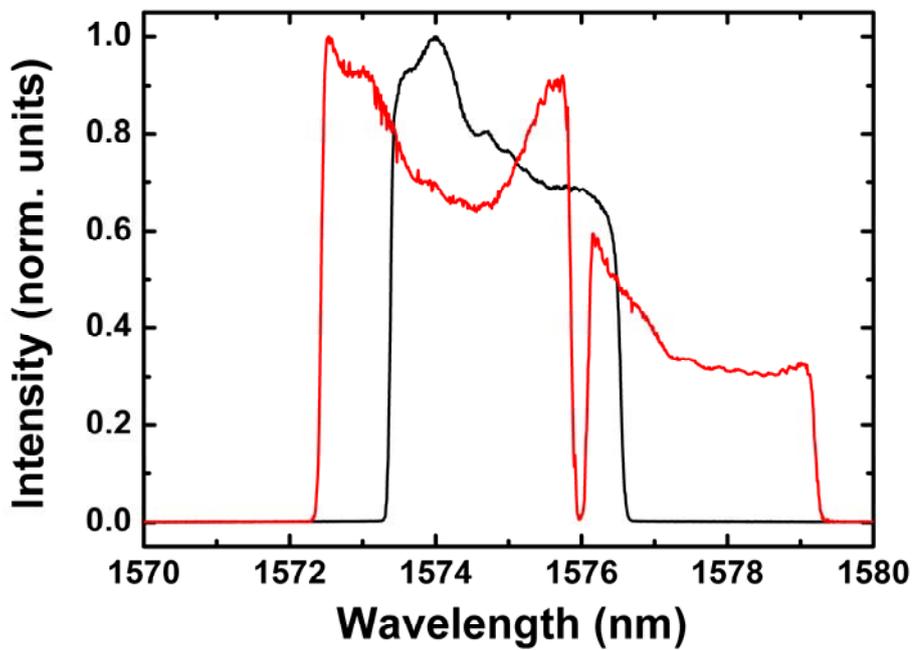

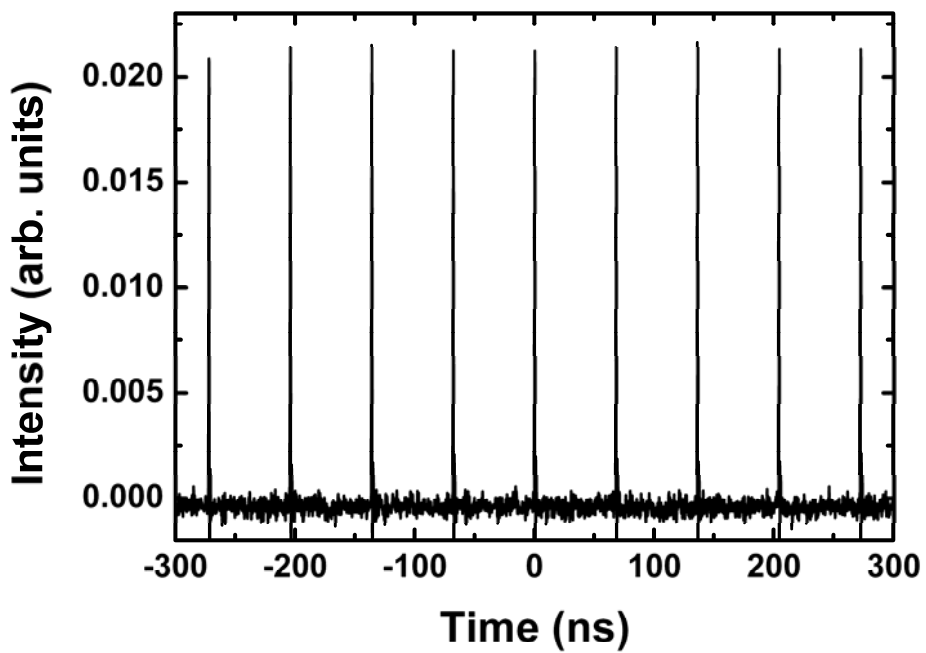

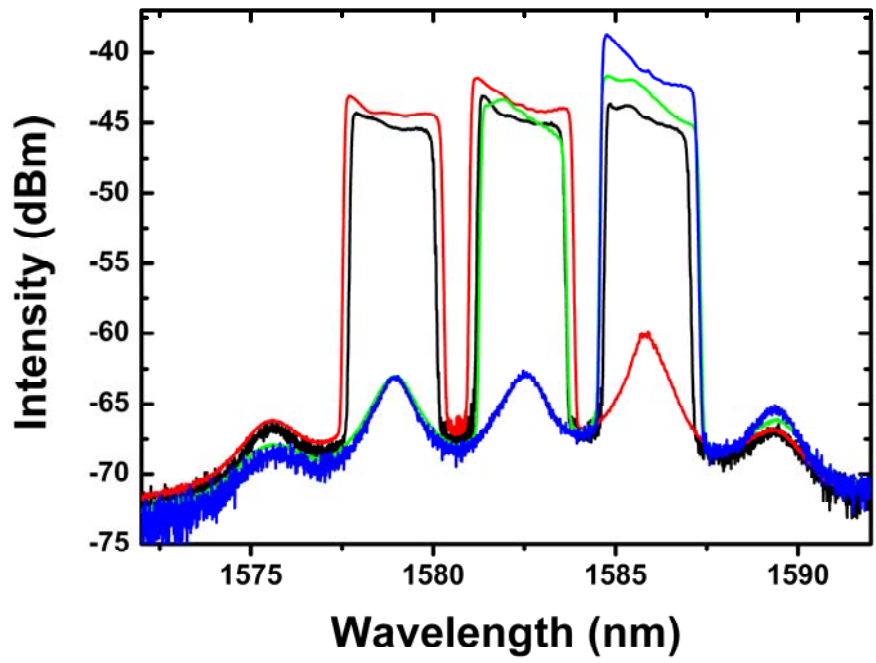
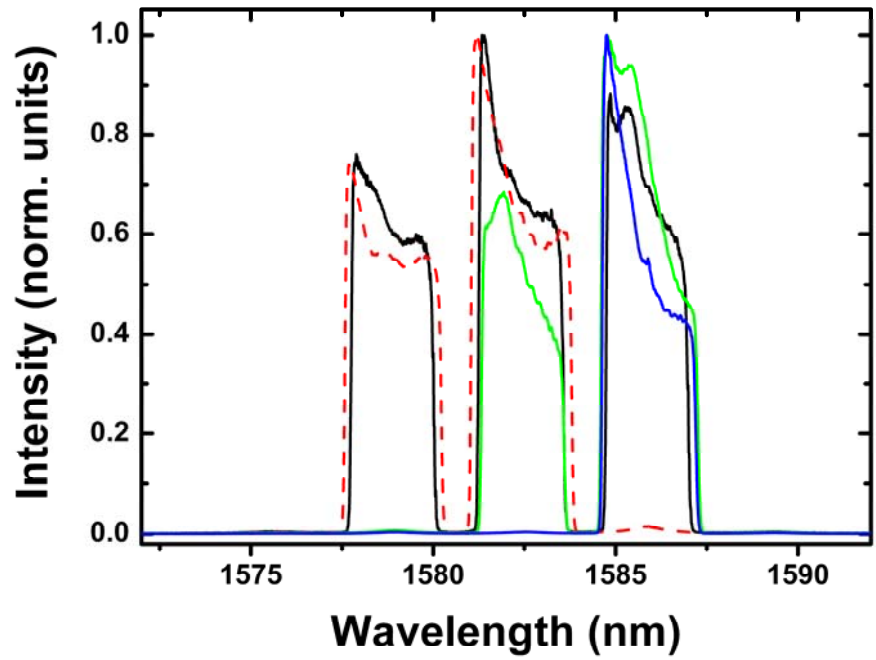